# A Novel Method for Image Integrity Authentication Based on Fixed Point Theory

Xu Li[a], Xingming Sun[b], Quansheng Liu[c], Beijing Chen[d]

[a] College of Information Science and Engineering, Hunan University, Changsha, 410082, China (e-mail: lixu_csust@ 126.com).

[b] College of Computer and Software, Nanjing University of Information Science & Technology, Nanjing, 210044, China (e-mail: sunnudt@163.com).

[c] Universit de Bretagne-Sud, Campus de Tohaninic, BP 573, 56017 Vannes, France (e-mail: quansheng.liu@univ-ubs.fr).

[d] College of Computer and Software, Nanjing University of Information Science & Technology, Nanjing, 210044, China (e-mail: cbj@nuist.edu.cn).

**Abstract:** Based on fixed point theory, this paper proposes a simple but efficient method for image integrity authentication, which is different from Digital Signature and Fragile Watermarking. By this method, any given image can be transformed into a fixed point of a well-chosen function, which can be constructed with periodic functions. The authentication can be realized due to the fragility of the fixed points. The experiments show that "Fixed Point Image" performs well in security, transparence, fragility and tampering localization.

## 1 Introduction

Among the current image integrity authentication techniques [1], Digital Signature [2] is the most wildly used method, because the asymmetric-key algorithm guarantees the security of Digital Signature. But this method is not suitable for tampering localization, and the additional signature information must be transmitted along with the original image. Fragile Watermarking [3] is the hottest research area in authentication. Its outstanding features are transparence and tampering localization, but its security is still a problem.

Recently, a new authentication scheme based on fixed point theory has been given in literature [4]. It has the same performance as Fragile Watermarking in tampering localization, but better performance in transparence. Since its authentication process can be considered a symmetric-key algorithm, we believe that its security is higher than most of Fragile Watermarking methods, but lower than Digital Signature methods. In the authentication scheme based on fixed point theory, the original image is transformed into a fixed point image by a well-chosen function, and the authentication is then implemented based on the fragility of the fixed point image. Some basic requirements were put forward in literature [4] for finding suitable functions, and a feasible function named Gaussian Convolution and Deconvolution (GCD) function was constructed in $\mathbf{Z}^{M \times N}$ (the $M \times N$ dimensional integer space). The fixed point images generated by a GCD function are very fragile, and have good localization ability for tampering. But unfortunately, there are few scattered and non-tampered points which may be wrongly marked as suspicious points.

In this paper, a new function is constructed in $\mathbf{Z}$ (the one dimensional integer space). This function has better performance in calculation, and its fixed point images can localize the tampering accurately in pixel-level. In fact, the localization result and the real tampering position differ with at most one pixel's distance.

## 2 Integrity authentication scheme based on fixed point theory

For a given mapping $f : D \to D$, if there exists $x \in D$ such that $f(x) = x$, then $x$ is called a fixed point of the mapping $f$ in the space $D$. In literature [4], $x$ is considered as an image, $f$ an operation on images, and $D$ a space of images; in order to make sure that $f$ is suitable for authentication, the author proposed three basic requirements: (1) for fragility, the function $f_k(\cdot)$ must have sparse fixed points for any given secret key $k$; (2) for calculation, any image can reach a fixed point of $f_k(\cdot)$ after several iterations; (3) for transparence, the original image and its fixed point image should not be distinguished visually.

For a fixed point image of $f$, if it is tampered, then after the calculation performed with $f$, the tampered image will move to a nearest fixed point. In general, these two fixed point images are not the same, so how to make the difference between them as small as we expect is a challenging problem. The GCD function can ensure that most of the false localizations happen near the tampered areas, and only few scattered and non-tampered points may be wrongly marked as suspicious points.

In order to make the localization more accurate, we discuss the integrity authentication problem in $D = \mathbf{Z}$ based on fixed point theory. Here, we consider $x$ as a single pixel of an image, $f$ an operation on pixels. This new function $f$ should satisfy three similar requirements for its fixed points: proper sparsity, feasibility of calculation, and good transparence.

Let's consider a simple function:

$$f(x) = x + R[\cos(x)], \qquad (1)$$

where

$$R(x) = \begin{cases} -1 & \text{if } x \leq -0.5, \\ 0 & \text{if } |x| < 0.5, \\ 1 & \text{if } x \geq 0.5. \end{cases}$$

Obviously, this function has fixed points in $\mathbf{Z}$, and these fixed points can be figured out easily. Specially, there are 86 uniformly dispersed fixed points in $\mathbf{Z}_{256} = \{0, 1, \cdots, 255\}$.

The function (1) is simple but not good enough for authentication in security, fragility and transparence. Taking into account the environment information, we construct the following new function. Let $I$ be a gray image of size $M \times N$, $x$ be the pixel value at $(s,t)$ of an image $I$, and $x_1, \cdots, x_8$ be the pixel values of the 8 closest neighbors of $(s,t)$ (see Fig. 1). Let $H$ and $K_1, \cdots, K_{12}$ be random matrices with the same size as $I$, the elements of $H$ taking values in $(0.5, 1]$ and the elements of $K_1, \cdots, K_{12}$ taking values in $[0, 2\pi]$. In addition, we suppose that at least half of $K_1, \cdots, K_8$ are equal to zero. For simplicity, let $h_{st} = H(s,t)$, $\mathbf{K}_{st} = (K_1(s,t), \cdots, K_{12}(s,t))$ and $\mathbf{X}_{st} = (x_1, \cdots, x_8, s, t, M, N)^T$, then for each position $(s,t)$, we define the function using for authentication as follows:

$$f_{\mathbf{X}_{st}, h_{st}, \mathbf{K}_{st}}(x) = x + R\left[h_{st} \cos(x + \mathbf{K}_{st}\mathbf{X}_{st})\right], \quad x \in \mathbb{Z}_{256}. \qquad (2)$$

We can give the functions for other pixels in the similar way. Specially, if there are some points that lie outside $I$, we specify the pixel values a constant, such as $0$.

| $x_1$ | $x_2$ | $x_3$ |
|---|---|---|
| $x_4$ | $x$ | $x_5$ |
| $x_6$ | $x_7$ | $x_8$ |

**Fig. 1.** The eight neighbors of the pixel $x$.

In $\mathbf{Z}_{256}$, the proportion of the fixed points of the function $f_{\mathbf{X}_{st},h_{st},\mathbf{K}_{st}}(\cdot)$ will decrease as $h_{st}$ increases, which results in bad quality of the fixed point images at the same time. Considering the transparence, we set the upper-bound of $h_{st}$ less than 1. The elements of the matrices $K_1,\cdots,K_{12}$ take random values in $[0,2\pi]$ because $\cos(\cdot)$ is a periodic function with period $2\pi$, and they are introduced to enhance the randomness of the fixed points of $f_{\mathbf{X}_{st},h_{st},\mathbf{K}_{st}}(\cdot)$. In addition, the parameters $s,t,M,N$ can effectively detect some copy attacks and geometric attacks.

For $x_1,\cdots,x_8$, we give the constraint that at least half of $K_1,\cdots,K_8$ are equal to zero. The reason is that if the number of non-zero parameters is less than 4, then $x$ and $x_1,\cdots,x_8$ will interact with each other in calculation, which bring difficulties for finding fixed point images. Under the constraint on $K_1,\cdots,K_8$, selecting proper way for point-by-point calculation can make the influences among $x$ and $x_1,\cdots,x_8$ to be one-way influences, which is very important for practical computation. For example, let $K_5,\cdots,K_8$ be equal to zero (the zero matrix), if we calculate from left to right and from top to bottom, then we will find in calculation process that the environments $x_1,\cdots,x_4$ can affect the generation of $\overline{x}$ (the corresponding fixed point of $x$), while $\overline{x}$ can affect the generation of $\overline{x}_5,\cdots,\overline{x}_8$ (the corresponding fixed points of $x_5,\cdots,x_8$). Similarly, if $K_1,\cdots,K_4$ are equal to zero, we can calculate from bottom to top and from right to left. Besides, at least one of $K_1,\cdots,K_8$ should be non-zero, which is for detecting collage attack.

In this model, the fragility can be considered as a probability problem. Suppose that $K_5,\cdots,K_8$ are equal to zero. If the fixed point $\overline{x}$ is tampered, then the modified value is no longer necessarily a fixed point; the same situation may occur for $\overline{x}_5,\cdots,\overline{x}_8$. For each position $(s,t)$, denote by $p_0$ the ratio of the number of fixed points of $f_{\mathbf{X}_{st},h_{st},\mathbf{K}_{st}}(\cdot)$ to 256, and by $p_1,\cdots,p_8$ the corresponding ratios for the 8 closest neighbors of $(s,t)$. Notice that $p_i$ ($i=0,1,\cdots,8$) is the probability for a pixel value to be a fixed point at the concerned position (the position $(s,t)$ or one of its 8 neighbors). If the fixed point $\bar{x}$ is tampered into $\bar{x}'$, the probability for $\bar{x}'$ and $\bar{x}_5,\cdots,\bar{x}_8$ to remain fixed points (at the corresponding positions) is $p_0 p_5 \cdots p_8$. Notice that although introducing more neighbors can make the fixed point images more fragile (cf. the analysis above), in order to locate tampering accurately, we only introduce eight neighbors $x_1,\cdots,x_8$. The strategy will be proven very satisfying by simulation.

## 3 Algorithm and Simulation

For a pixel value $x_0$ at position $(s,t)$ of an image $I$, we can find a fixed point of $f_{\mathbf{X}_{st},h_{st},\mathbf{K}_{st}}(\cdot)$ after several iterations, but the fixed point may not be the nearest fixed point to $x_0$, and it may jump out of $\mathbb{Z}_{256}$. Considering the transparence and the value range of gray images, we transform the iteration problem into a problem of integer optimization as follows:

$$\begin{cases} \min(|x-x_0|) \\ s.t. \quad |h_{st}\cos(x+\mathbf{K}_{st}\mathbf{X}_{st})|<0.5 \\ \quad\quad x\in[x_0-3,x_0+3] \\ \quad\quad x\in\mathbb{Z}_{256} \end{cases} \quad (3)$$

The constraint $x\in[x_0-3,x_0+3]$ is introduced to simplify the optimization calculation, so that we can get the suitable fixed point by exhaustive search algorithm. We choose $[x_0-3,x_0+3]$ because at most 3 integers can satisfy $|\cos(\cdot)|<0.5$ in an interval of length $2\pi/3$, which means that the maximal distance between $x_0$ and the fixed points of $f_{\mathbf{X}_{st},h_{st},\mathbf{K}_{st}}(\cdot)$ is no more than 3.

### 3.1 Algorithm.

1) The sender and the receiver establish the secret key.

2) For a given image $I$, the sender generates $H$ and $K_1,\cdots,K_{12}$ with the secret key, and calculates the fixed point of every pixel to get the fixed point image $J$, and then send $J$ to the receiver over a public channel.

3) For the received image $J'$, the receiver generates parameters similarly with the secret key, and verifies whether $|h_{st}\cos(y+\mathbf{K}_{st}\mathbf{Y}_{st})|<0.5$ for all $y\in J'$, for each position $(s,t)$. If the inequality fails for some points in $J'$, we mark them to be suspicious points;

otherwise we conclude that $J' = J$.

**3.2 Key generation scheme and Security.**

For each pixel value $x$ at position $(s,t)$ of an image $I$, if $\bar{x}$ is a fixed point corresponding to $x$, it is impossible to figure out $h_{st}$ and $\mathbf{K}_{st}$ because of the truncating operation on $\cos(\cdot)$. So the above algorithm can be considered as a kind of symmetric key algorithm, and the security depends only on the size of the key space.

The key generation scheme is based on a random number generator. We choose the following linear congruential generator [5] to produce random numbers:

$$x_{n+1} = \left[(4a+1)x_n + 2b+1\right] \mod(2^m), \quad n \geq 0, \qquad (4)$$

where $x_0, a, b$, and $m$ take positive integers. The random sequence $x_0, x_1, \cdots$ can be standardized into a uniform distribution on any given interval, so that a random matrix in any interval can be generated with a set of parameters $(x_0, a, b, m)$. Then the key can be constructed with 13 sets of parameters $\{(x_{0i}, a_i, b_i, m_i) : i = 1, \cdots, 13\}$, and the key space is big enough for practical use if these parameters are not too small. For example, if the elements of the key take random integers in $[10, 50]$, the key space is larger than $41^{4 \times 9} \approx 2^{192}$.

**3.3 Transparence.**

The Peak Sign-to-Noise Ratio (PSNR) is selected as the similarity measure between an image and its fixed point images; 2000 images are selected for simulation from the image database FREEFOTO [6]. Fig. 2 shows the simulation result, where the elements of the key take random integers in $[10, 90]$ and the quality of the fixed point images is adjusted via the upper bound of $h_{st} = H(s,t)$.

Compared with the experiment results in literature [4], the fixed point images in this paper perform less well in transparence. The reason is that the sparsity of the fixed points is harder to control in $\mathbf{Z}$ than that in $\mathbf{Z}^{M \times N}$. Happily, both methods get higher PSNR value than most of the fragile or semi-fragile watermarking methods; when the upper bound of $H(s,t)$ is less 0.7, even the minimum value of PSNR is larger than 51 dB. Additionally, the PSNR value can be adjusted freely in a continuous interval in both methods.

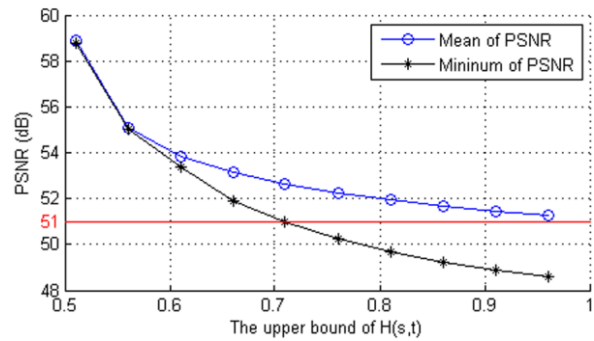

**Fig. 2.** Test for transparence.

**3.4 Fragility.**

By the discussion in Section 2, we know that fixed point images can detect the common attacks such as image processing, JPEG compression, etc., because these attacks significantly modify the fixed point images and these modified images are almost impossible to pass the integrity authentication. In Fig. 3, the Rewriting attack experiment demonstrates the fragility of a fixed point image and the security of our scheme at the same time. The image of size 400×600 is selected from the image database FREEFOTO; the upper bound of $H(s,t)$ is equal to 0.52; the elements of the key take random integers in $[10, 90]$; the suspicious pixels are marked with white dots.

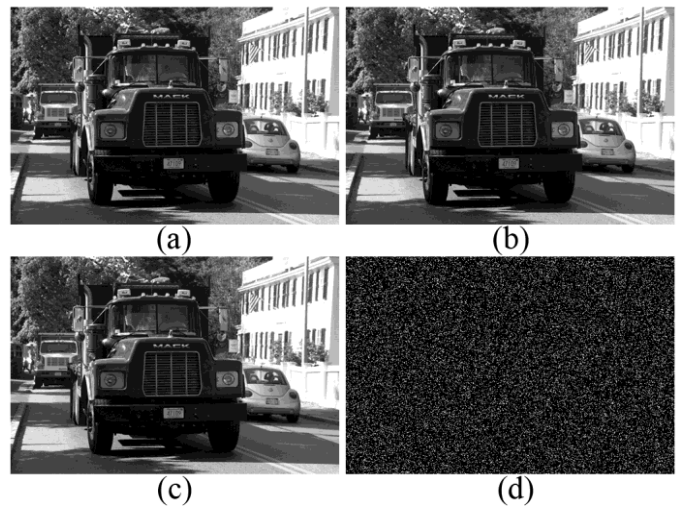

**Fig. 3.** Test for fragility security. (a) is the original image; (b) is the fixed point image with PSNR=57.3503 dB; (c) is the Rewriting attack with known original image, where the upper bound of $H(s,t)$ is equal to 0.51, and PSNR=58.8717 dB; (d) is the corresponding authentication results.

**3.5 Tampering localization.**

Using the same original image, we test the tampering localization ability against some common local attacks, where the tampered areas are marked with rectangles. Fig. 4 shows the comparison of our tampering localization results (Fig.4 (c, d)) and the result of the method in literature [4] (Fig.4 (b)). We see that the fixed point images in this paper perform better in tampering localization as we discussed in Section 2. In addition, the attack responses (Fig.4 (d)) are stronger when the upper bound of $H(s,t)$ is higher.

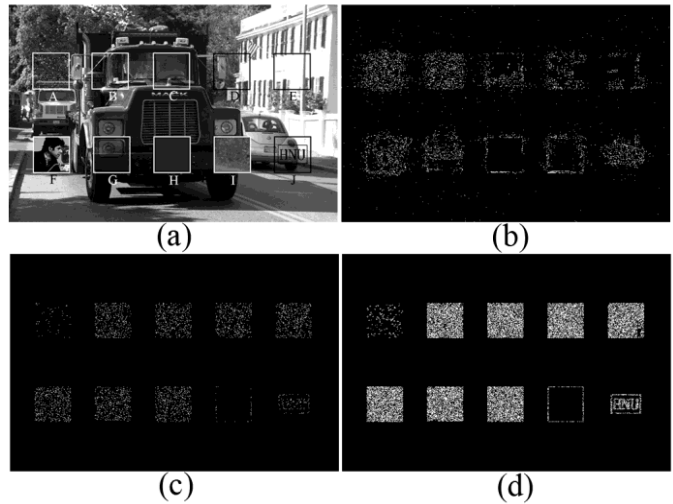

**Fig. 4.** Test for tampering localization. (a) shows some common localized modifications, where A is the Salt and Pepper noise attack; B is the Gaussian noise attack; C is the Median filter attack; D is the Gaussian filter attack; E is the Image enhancement attack; F is the Copy attack from an another image; G is the Copy attack from itself; H is the Covering attack with a constant; I is the Collage attack from an another fixed point image generated with the same parameters; J is adding illegal logo; (b) shows the authentication result of the method in literature [4], and the PSNR value of the fixed point image is equal to 58.5374 dB; (c) shows the result of the method in this paper, where the upper bound of $H(s,t)$ is equal to 0.512, and PSNR=58.4268 dB; (d) shows the result when the upper bound of $H(s,t)$ is equal to 1, and PSNR=50.9151 dB.

**Note.** For the Collage attack in the area I (Fig.4 (a)), the authentication results (Fig.4 (c, d)) show up as hollow squares, which are caused by the selecting method of environment information in (2). It is worth noticing that this phenomenon can be used to distinguish the Collage attack from other

attacks.

## 4 Conclusions

We have proposed a new method for image integrity authentication, based on fixed point theory. The proposed method can deal with gray images with simple algorithm, fast calculation, high security and high-precision localization. We find that it is more suitable for the "Trustworthy Digital Camera" since it can efficiently treat a lot of images and it can also be easily generalized to color image authentication.

**Acknowledgments**

This work is supported by the NSFC (61232016, 61173141, 61173142, 61173136, 61103215, 61070196, 61070195, and 61073191), National Basic Research Program 973 (2011CB311808), 2011GK2009, GYHY201206033, 201301030, 2013DFG12860 and PAPD fund.